# Electron mobility in few-layer $Mo_xW_{1-x}S_2$

Hareesh Chandrasekar and Digbijoy N. Nath[,a)]

Centre for Nano Science and Engineering,
Indian Institute of Science, Bangalore 560012 (India)

**Abstract**: In this letter, we theoretically study the electron mobility in few-layer $Mo_xW_{1-x}S_2$ as limited by various scattering mechanisms. The room temperature energy-dependent scattering times corresponding to polar longitudinal optical (LO) phonon, alloy and background impurity scattering mechanisms are estimated based on the Born approximation to Fermi's Golden rule. The contribution of individual scattering rates is analyzed as a function of 2D electron density as well as of alloy composition in $Mo_xW_{1-x}S_2$. While impurity scattering limits the mobility for low carrier density ($<2 \times 10^{12}$ cm$^{-2}$), LO polar phonon scattering is the dominant mechanism for high electron densities. Alloy scattering is found to play a non-negligible role for $0.5 < x < 0.7$ in $Mo_xW_{1-x}S_2$. The LO phonon limited and impurity limited mobilities show opposing trends with respect to alloy mole fraction. The understanding of electron mobility in $Mo_xW_{1-x}S_2$ presented here is expected to aid the design and realization of heterostructures and devices based on alloys of $MoS_2$ and $WS_2$.

---

a) Author to whom correspondence should be addressed.
   Electronic mail: digbijoy@cense.iisc.ernet.in

Current research on layered 2D transition metal dichalcogenides (TMDs) is largely motivated by the wide range of physical phenomena they exhibit, the ability to achieve ultra-high confinement[1] which is not possible with bulk semiconductors, and the absence of lattice mismatch issue for heteroepitaxial integration on any substrate.[2] These advantages combined with the sheer divergence of their electronic and opto-electronic properties[3] make them highly promising for a wide range of next generation device applications otherwise not viable with traditional semiconductors. Among these materials $MoS_2$[4,5] and to a lesser extent $WS_2$[6] have been extensively explored as promising semiconductor counterparts to 2D metallic conductors and insulators such as graphene and boron nitride respectively, and have been touted as potential channel materials to replace silicon based technologies and for flexible electronics. In addition to the individual materials themselves, heterostructures of these 2D materials are being increasingly explored in order to exploit and extend their functionalities in unique ways. One of the key underpinnings of any heterostructure based technology is the ability to engineer alloys in a controllable fashion preferably spanning the entire range of possible compositions. Such alloy engineering facilitates not only band gap tunability, but also control over other electronic properties such as band offsets, type of charge carriers and effective masses all of which are crucial for optimal heterostructure design.

In view of the prospects held out by $MoS_2$, $WS_2$ and their lateral and vertical heterojunctions for electronics,[7] alloys of $MoS_2$-$WS_2$ are of particular interest. The similarity between their crystal structures and near identical lattice parameters contribute to making layered $Mo_xW_{1-x}S_2$ thermodynamically stable in their both their bulk and 2D forms.[8,9] $Mo_xW_{1-x}S_2$ layers have been physically realized using solution synthesis,[8] exfoliation from chemical vapor transport grown single crystals,[10,11] chemical vapor deposition[12] and sulphurization of co-sputtered Mo and W films.[13] In-plane composition variations across the full range from $MoS_2$ to $WS_2$ have also been demonstrated.[13,14] Single layered $Mo_xW_{1-x}S_2$ has also been shown to form a perfectly random alloy across the entire range of compositions[11] which is desirable for use in heterostructures. Band gap tunability over the full composition range has been observed[10] and the possibility of using 2D $Mo_xW_{1-x}S_2/TiO_2$ heterostructures for photo-catalytic applications and water-splitting due to band engineering at the interface has also been proposed.[15] Carrier injection using high work function metals into $MoS_2$ is most commonly observed to be n-type and is expected to be p-type in $WS_2$ due to the nature of their band offsets, and hence alloy engineering has been predicted as a possible means of changing the type of carriers from electrons to holes and also offers a means to alter the effective masses of both carriers.[16] However, despite the relevance of $Mo_xW_{1-x}S_2$ to device applications, studies of carrier transport through such alloy layers have been scarce. Few layer (FL) and multi-layer (ML) TMDs in particular have been shown to carry higher current densities and are promising for practical device applications.[5,17] In this letter, we study electron mobility in FL $Mo_xW_{1-x}S_2$ by estimating the energy-dependent scattering time using a 2D formalism as a function of both carrier density and alloy composition.

The structure employed in this study is shown in Fig. 1(a). We consider a 5 nm thick $Mo_xW_{1-x}S_2$ on a sapphire substrate with $Al_2O_3$ as the top-gate dielectric. The lateral transport in such FL TMD is 2D in nature because carriers move independently in each layer[5] and the interlayer interaction can be neglected. Thus, 2D transport formalism is applicable for understanding transport in FL $Mo_xW_{1-x}S_2$. The choice of $Al_2O_3$ is guided both by its extensive use as a gate dielectric for top gated transistors, and by its efficacy as evidenced by recent results[5,17] reporting an enhancement of carrier mobility in ML $MoS_2$ with the use of $Al_2O_3$.

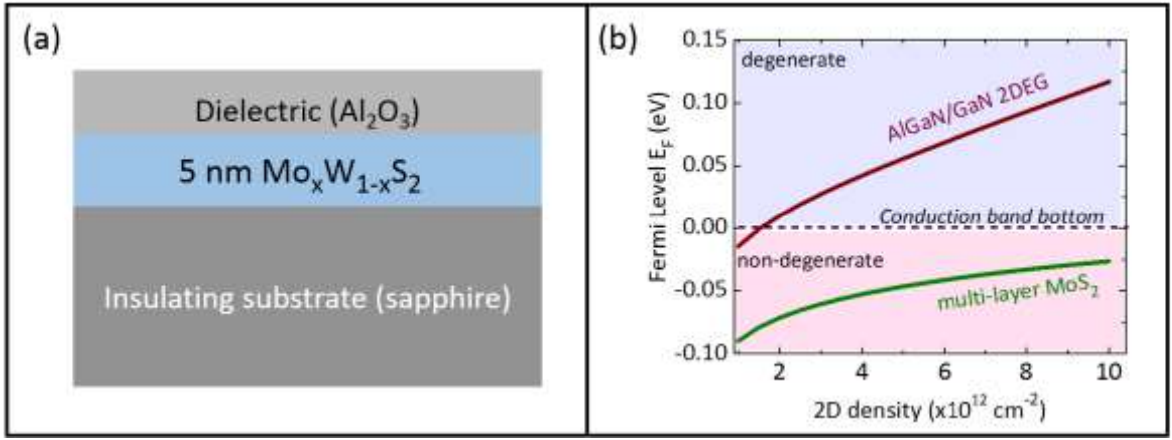

**Fig. 1:** (a) The schematic of the structure considered in this study (b) Fermi level as a function of 2D carrier density for $MoS_2$ and for the polar AlGaN/GaN system – a wide band gap transistor, highlighting the non-degeneracy of $MoS_2$ even with an electron density ~ $10^{13}$ $cm^{-2}$.

The analysis of carrier transport in a 2D electron gas depends critically on whether or not the system is degenerate. In 2D transport such as in conventional AlGaAs/GaAs or AlGaN/GaN, the electron gas is degenerate when the Fermi level lies above the conduction band in the notch formed at the heterojunction. If the bottom of the conduction band is taken as the reference (i.e. zero), then the position of the Fermi level in 2D systems is given by:

$$E_F = kT \log\left(\exp\left(\frac{\pi \hbar^2 n_{2D}}{g_v \, m \, kT}\right) - 1\right) \qquad (1)$$

Here, $n_{2D}$ is the carrier density ($cm^{-2}$), $g_v$ is the valley degeneracy and m is the effective mass of carriers in the material. Bulk $MoS_2$ has a high in-plane effective mass of around $0.71m_0$ (where $m_0$ is the electron mass) which is extracted based on the average in-plane mass in transverse and longitudinal directions as calculated by Density Functional theory (DFT).[18] The valley degeneracy for bulk $MoS_2$ is $g_v = 6$.[5] As shown in Fig. 1(b), the Fermi level in bulk or multi-layer $MoS_2$ lies below the conduction band bottom even for an otherwise high carrier density of $10^{13}$ $cm^{-2}$. This implies that the *2D electron gas in $MoS_2$ is non-degenerate even up to $10^{13}$ $cm^{-2}$*, as opposed to that in AlGaN/GaN system (Fig. 1(b)) which consists of a degenerate electron gas for $n_{2D} > 10^{12}$ $cm^{-2}$. The in-plane effective mass of electrons in bulk

WS$_2$ is 0.62m$_0$ with the same g$_v$ = 6 as in MoS$_2$.[18] Thus, the 2D electron gas in WS$_2$ is also non-degenerate for carrier densities considered in this study. As such, the scattering time has to be evaluated for non-degenerate electron gas in the alloy Mo$_x$W$_{1-x}$S$_2$, implying that we cannot consider the electron transport to be occurring at the Fermi level unlike in a degenerate electron gas.[19]

The scattering mechanisms considered for electron transport in Mo$_x$W$_{1-x}$S$_2$ are scattering due to (i) alloy (ii) polar LO phonon (iii) background 3D impurities. It has been shown[5,17] that other scattering processes such as acoustic phonons and homopolar optical phonons have little contribution to carrier mobility in multi-layer MoS$_2$ and are hence not considered here. The various material parameters for Mo$_x$W$_{1-x}$S$_2$ such as effective mass, in-plane and out-of-plane lattice constants and static and high-frequency dielectric constants are linearly interpolated between the values of MoS$_2$ and WS$_2$ without the use of any bowing parameter. Table I provides the material parameter values for MoS$_2$ and WS$_2$ used for this study.

|  | MoS$_2$ | WS$_2$ |
|---|---|---|
| Effective Mass (mean of m$_l$ and m$_t$) | 0.71m$_0$[18] | 0.62m$_0$[18] |
| Lattice parameters – a,c (Å) | 3.127, 12.066[20] | 3.126, 12.145[20] |
| Static and high frequency dielectric constants – $\varepsilon_s$, $\varepsilon_\infty$ | 7.6, 7.0[5] | 7.0,[21] 5.76[22] |

**Table I:** Mean effective masses, lattice parameters, and static and high frequency dielectric constants of MoS$_2$ and WS$_2$ used in this study. The high-frequency dielectric constant for WS$_2$ is taken for 3R- WS$_2$ due to non-availability of data for 2H-WS$_2$. However given the structural similarities between the two, the value is expected to be fairly close. It should be noted that the corresponding values for monolayer 2H-MoS$_2$ and 3R-MoS$_2$ for instance, are almost identical.[23]

For alloy scattering in a 2D system such as in AlGaN/GaN, the electron gas exists in the binary compound (i.e. GaN) with the tail of the electron wave-function penetrating in to the ternary AlGaN which gives rise to alloy scattering. Typically, a modified Fang-Howard wave-function is used to model the penetration in to the ternary alloy barrier. The more the wave-function penetration or more disordered the alloy is, the lower would be the carrier mobility. However in the present structure under zero bias, we can safely assume the FL $Mo_xW_{1-x}S_2$ to be a quantum well sandwiched between two infinitely high potential barriers, which is justified because of the large band gap (and hence conduction band discontinuity $\Delta E_C$) of $Al_2O_3$ compared to $Mo_xW_{1-x}S_2$. As a result, we can assume that there is no penetration of electron wave-function in to the dielectric as well as in to the sapphire substrate.

The momentum scattering time ($\tau_{alloy}$) in a 2D electron gas due to alloy disorder is obtained from applying the Born approximation to Fermi's Golden rule as[19]

$$\frac{1}{\tau_{alloy}(x)} = m(x)\frac{\Omega(x)}{\hbar^3}[\delta V]^2 x(1-x)\int[\psi(z)]^4 dz \qquad (2)$$

where, $\Omega(x)$ is the volume of the unit wurtzite cell[24] of $Mo_xW_{1-x}S_2$ and $\delta V$ is the fluctuating potential which can be taken to conduction band offset $\Delta E_C = 0.27$ eV between $MoS_2$ and $WS_2$.[25] The electron wave-function in the $Mo_xW_{1-x}S_2$ can be taken as $\psi(z) = \sqrt{\frac{2}{d}}\sin\left(\frac{\pi z}{d}\right)$, (where d = thickness of $Mo_xW_{1-x}S_2$) corresponding to that of an infinite square well as justified earlier. Further, the change in 2D electron density is assumed to be achieved by varying the gate bias which implies the field inside the quantum well changes but to a first order, we shall stick to an infinite potential well model for simplification of analysis. Interestingly therefore, the alloy scattering does not depend on the density of the electron gas or on background impurity and instead has a linear dependence on the thickness of the layer.

The polar LO phonon energy for MoS$_2$ is $E_{op}$(MoS$_2$) = 48 meV[5]. The phonon dispersion relation for bulk WS$_2$ yields a value of 360 cm$^{-1}$ at the high symmetry $\Gamma$ point for the IR-sensitive LO branch which translates to $E_{op}$(WS$_2$) = 44.6 meV[20]. The highest attainable kinetic energy by 2D electrons in the non-degenerate carrier system is at the Fermi level and is given by $E_F = \hbar^2 k_F^2/(2m)$, where $k_F = \sqrt{(2\pi n_{2D})}$ is the Fermi wave-vector. Even for $n_{2D} = 10^{13}$ cm$^{-2}$, the carrier kinetic energies for MoS$_2$ and WS2 are 33 meV and 38.7 meV respectively, which are both less than their respective optical phonon energy. This combined with the fact that the $E_{op}$ for both MoS$_2$ and WS$_2$ is larger than the thermal energy of carriers at room temperature ($\hbar\omega_{op} > k_B T$), imply that most carriers have energies lower than $E_{op}$, thus blocking phonon emission[26]. Thus, phonon absorption process will dominate and the momentum relaxation rate of electrons in an infinite potential well for phonon absorption based on Fermi's Golden rule is given by[27]:

$$\frac{1}{\tau_{LO}} = \frac{e^2 \omega_{op} n(\omega_{op}) m^* d}{2\varepsilon_p \hbar^2} \qquad (3)$$

Here, $\omega_{op}$ is the phonon frequency, $n(\omega_{op})$ is the number of phonons given by Bose-Einstein statistics as:

$$n(\omega_{op}) = e^{\frac{\hbar\omega_{op}}{2\pi k_B T}} - 1 \qquad (4)$$

$\varepsilon_p^{-1} = \varepsilon_\infty^{-1} - \varepsilon_s^{-1}$ where $\varepsilon_\infty$ and $\varepsilon_s$ are respectively the static and high frequency dielectric constants which for Mo$_x$W$_{1-x}$S$_2$ are linearly interpolated between the values of the two binary compounds as mentioned earlier.

It is interesting to note here that the LO phonon scattering rate *increases* linearly with thickness of the layers 'd', implying that the LO phonon limited mobility will degrade with increasing thickness of Mo$_x$W$_{1-x}$S$_2$. It needs to be pointed out that a *reduction* in LO polar phonon scattering rate with increasing thickness of $MoS_2$ has been estimated for multilayers of $MoS_2$, described using a Fang-Howard formalism[28], contrary to our prediction. This is primarily due to the thickness dependence in the Fang-Howard wavefunction itself used in ref. [28]. It has also been reported[29] that the LO phonon limited mobility is fairly independent of carrier density up to ~ $10^{13}$ cm$^{-2}$, and a weak dependence is observed when screening is considered. Equation (3) also indicates that it is independent of carrier density but depends on alloy fraction 'x' through the effective mass and dielectric constant dependencies.

To estimate background impurity scattering on 2D carrier transport, we consider the expression for an arbitrary wave-vector 'k' corresponding to energy E, as:[19]

$$\frac{1}{\tau_{imp}(n_{2D},x,E)} = n^{3D}_{imp} \frac{g_v\, m}{2\pi \hbar^3 k(E)^3} \left(\frac{e^2}{2\varepsilon(x)}\right)^2 \int \left[\frac{1}{q+q_{TF}(x)}\right]^2 \frac{q}{\sqrt{1-\left(\frac{q}{2k(E)}\right)^2}} dq \quad (5)$$

The wave-vector and energy share the relation $k(E) = \sqrt{2mE/\hbar^2}$. Although the electron gas is non-degenerate, we nevertheless consider the effect of screening of impurities by carriers due to their high density. So we use $q_{TF}$ as the Thomas-Fermi screening wave-vector defined as:[19]

$$q_{TF}(x) = g_v \frac{m(x)}{2\pi \hbar^2} \frac{e^2}{\varepsilon(x)} \frac{\partial n_{2D}(E)}{\partial E_F} \quad (6)$$

The energy-dependent relaxation time as given by equation (6) is now averaged over all energies (till Fermi level) using a Fermi-Dirac distribution[29]:

$$\frac{1}{\tau_{imp}(n_{2D},x)} = \frac{1}{n_{2D}} \int_{-\infty}^{E_F} \frac{1}{\tau_{imp}(n_{2D},x,E)} \frac{g_v\, m}{\pi \hbar^2} E\left(-\frac{\partial f(E, E_F)}{\partial E}\right) dE \quad (7)$$

In our calculations, we shall assume a typical background impurity concentration of $10^{17}$ cm$^{-3}$ corresponding to $5\times10^{10}$ cm$^{-2}$ of equivalent 2D impurity density.

The resulting room temperature electron mobility in $Mo_xW_{1-x}S_2$ limited by alloy and polar LO phonon scattering as a function of alloy composition is shown in Fig. 2(a). We see that, as expected, mobility due to alloy scattering decreases with an increase in alloy fraction reaching a minimum at x=0.54. The difference in the effective masses of $MoS_2$ and $WS_2$ gives rise to the asymmetry in the mobility about x=0.54 (inset of Fig. 2(a)), the lighter effective mass of $WS_2$ giving rise to a slightly higher mobility in the W-rich alloy compositions when compared to the same fraction of Mo-rich alloys. We also note that although it plays a non-negligible role in the total mobility particularly in the $0.5 < x < 0.7$ range, yet alloy scattering is the limiting scattering mechanism at 300 K. The carrier density dependent background impurity limited mobility is plotted alongside the polar LO phonon mobility in Fig. 2(b) for two carrier densities of $1\times10^{12}$ cm$^{-2}$ and $2\times10^{12}$ cm$^{-2}$ and a nominal background impurity concentration of $1\times10^{17}$ cm$^{-3}$ for the entire range of alloy compositions at 300 K.

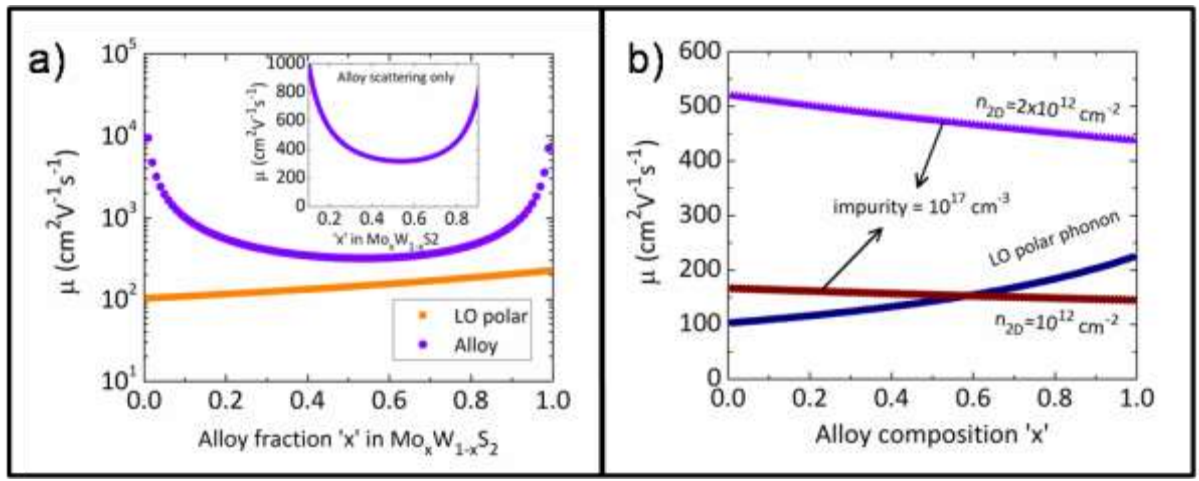

**Fig. 2:** (a) Electron mobility (log scale) due to polar optical phonon scattering and alloy scattering as a function of alloy mole fraction 'x'. Inset shows the contribution of alloy scattering alone with composition (linear scale). (b) Mobility due to impurity scattering for a background impurity density of $10^{17}$ cm$^{-3}$ compared with the LO phonon limited mobility at two different carrier densities – $1\times10^{12}$ cm$^{-2}$ and $2\times10^{12}$ cm$^{-2}$ - as a function of alloy composition.

It is observed that these two scattering mechanisms (i.e. background impurity and LO phonon) display contrasting trends as a function of alloy composition, with the impurity scattering mobility decreasing with increasing alloy fraction whereas LO phonon mobility increases. The decrease in $\mu_{imp}$ with alloy composition is due to the higher effective mass of $MoS_2$ as compared to $WS_2$ while $\mu_{LO}$ increases with increasing Mo-fraction resulting from the differences in the static and high frequency dielectric constants. It is evident that the electron mobility is limited only by the LO phonon scattering process across the entire alloy fraction for carrier densities of $2x10^{12}$ cm$^{-2}$ and above while this is not the case for lower carrier densities. For example, at a carrier density of $1x10^{12}$ cm$^{-2}$ the polar optical phonon limits the mobility in the W-rich regions (x≤0.5) while it is the impurity scattering that becomes the dominant scattering mechanism with increasing Mo fraction.

We now examine the relative effects of the three scattering mechanisms on the total mobility of electrons using Matthiessen's rule ($1/\mu_{tot}=1/\mu_{LO}+1/\mu_{imp}+1/\mu_{alloy}$). Fig. 3 shows the total electron mobility along with the mobilities due to the individual scattering mechanisms at a nominal alloy composition of x=0.6 for various 2D carrier densities at 300 K. It is clear that the total mobility is primarily limited by LO phonon scattering although the role of alloy scattering is not insignificant. Impurity scattering only plays a role at low carrier densities as discussed above. It should be noted that the effect of impurity scattering on total mobility will increase if the concentration of impurities is increased from $10^{17}$ cm$^{-3}$ to higher values while impurity concentrations $<10^{17}$ cm$^{-3}$ will have lesser impact on the overall mobility even at low 2D carrier densities.

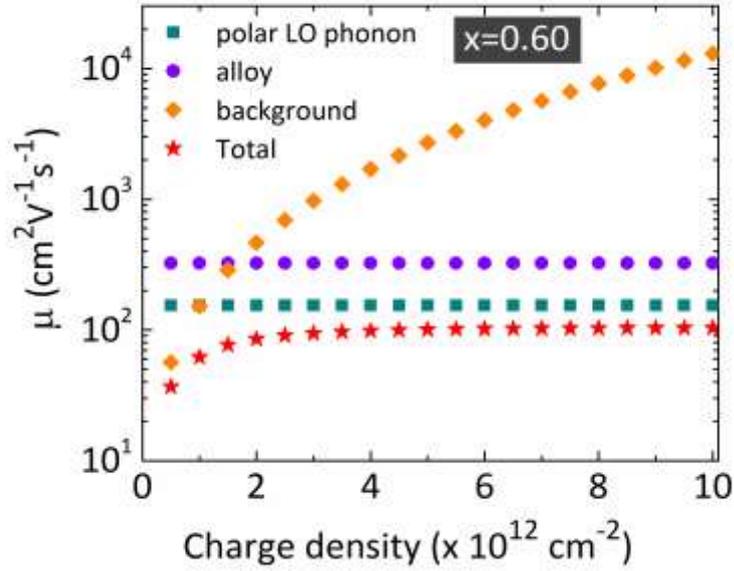

**Fig. 3:** Electron mobility due to the individual alloy, LO phonon, and background impurity scattering mechanisms and the resulting total mobility as a function of 2D carrier density for an alloy composition of x=0.6 and impurity density of $10^{17}$ cm$^{-3}$.

To enable the appropriate choice of alloy composition for device applications, we now present the total electron mobility for some technologically important carrier densities across the entire range of alloy compositions in Fig 4. We see that the total electron mobility is pegged to around 100 cm$^2$/Vs for $WS_2$ of 5 nm thickness at moderate carrier densities of $5\times10^{12}$ cm$^{-2}$ primarily due to LO phonon scattering. Increasing the Mo-fraction in the alloy increases the mobility until it reaches ~250 cm$^2$/Vs corresponding to few-layer $MoS_2$. Experimental values of electron mobility for few layer (8-10 nm) $WS_2$[30] and $MoS_2$[31] indicated on the plot fall within the range of our predictions and also display the trend of increasing mobility from $WS_2$ to $MoS_2$. This clearly indicates the desirability of increasing the alloy fraction 'x' to raise mobility and hence improve device performance provided the band gap and band offset requirements are met.

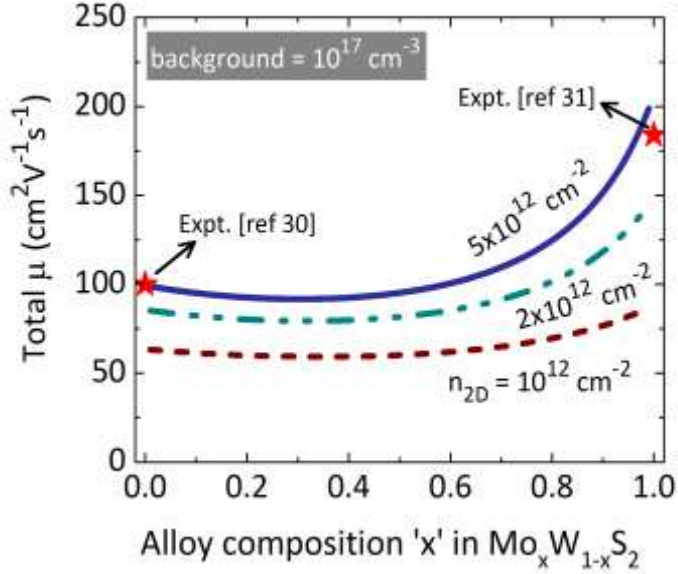

**Fig. 4:** Total electron mobility as a function of alloy composition 'x' for 2D carrier densities of $1\times10^{12}\,cm^{-2}$, $2\times10^{12}\,cm^{-2}$ and $5\times10^{12}\,cm^{-2}$ for a background impurity density of $10^{17}\,cm^{-3}$ at 300K. The stars indicate experimentally reported values[30,31] of mobility for multilayer (8-10 nm) $WS_2$ and $MoS_2$ shown for comparison.

At lower carrier densities, impurity scattering further reduces both the mobility and also the variation in mobility with alloy composition. Hence we see that unlike monolayer $WS_2$, few-layer $WS_2$ and hence W-rich alloys of $Mo_xW_{1-x}S_2$ display poorer electron mobility and hence lower performance as a channel material for transistors as compared to Mo-rich compositions.

In conclusion, the mobility of few-layer $Mo_xW_{1-x}S_2$ films at room temperature has been investigated considering the contributions of polar optical phonons, alloy scattering and background impurity scattering mechanisms under the framework of Fermi's Golden Rule. While alloy scattering was found to be important for alloy compositions $0.5<x<0.7$, the mobility is found to be limited by LO phonon scattering across the entire range of compositions except for the combination of Mo-rich alloys with very low 2D carrier densities ($\leq10^{12}\,cm^{-2}$) where the impurity scattering dominates. The intrinsic electron mobility of the alloy is found to increase with increasing the Mo-fraction primarily due to its lower LO phonon scattering rates as compared to the same fraction of W-rich alloy. This makes the use of Mo-rich $Mo_xW_{1-x}S_2$ layers more desirable for a given band gap-band offset requirement in designing heterostructures. This study can be further extended to holes as the pre-dominant carriers and for biased/gated layers using a Fang-Howard or modified Fang-Howard formalism to take into account the modification of the electron wave-function due to applied bias, as also for investigating transport in other alloy systems such as $MSe_2$, $MTe_2$, $MS_xSe_{1-x}$ (where M=Mo,W). We believe that the results presented here will enable the appropriate selection of TMD based alloys for use either individually or in hetero-structures for applications ranging from electronics and opto-electronics to photocatalysis.

**Acknowledgment:** The authors would like to acknowledge helpful discussions with Dr. Navakanta Bhat and Dr. Srinivasan Raghavan. Digbijoy Nath acknowledges the funding from XII plan start-up grant of IISc [R(VI)090/978/2012-751].

# References

1   L. T. Liu, S. B. Kumar, Y. Ouyang, and J. Guo,  Ieee Transactions on Electron Devices 58 (9), 3042 (2011).
2   A. K. Geim and I. V. Grigorieva,  Nature 499 (7459), 419 (2013).
3   J. A. Wilson and A. D. Yoffe,  Advances in Physics 18 (73), 193 (1969).
4   B. Radisavljevic, A. Radenovic, J. Brivio, V. Giacometti, and A. Kis,  Nature nanotechnology 6 (3), 147 (2011);   D. Krasnozhon, D. Lembke, C. Nyffeler, Y. Leblebici, and A. Kis,  Nano letters 14 (10), 5905 (2014).
5   S. Kim, A. Konar, W. S. Hwang, J. H. Lee, J. Lee, J. Yang, C. Jung, H. Kim, J. B. Yoo, J. Y. Choi, Y. W. Jin, S. Y. Lee, D. Jena, W. Choi, and K. Kim,  Nature communications 3, 1011 (2012).
6   D. Braga, I. Gutierrez Lezama, H. Berger, and A. F. Morpurgo,  Nano letters 12 (10), 5218 (2012);   W. S. Hwang, M. Remskar, R. S. Yan, V. Protasenko, K. Tahy, S. D. Chae, P. Zhao, A. Konar, H. L. Xing, A. Seabaugh, and D. Jena,  Applied Physics Letters 101 (1) (2012).
7   Y. J. Gong, J. H. Lin, X. L. Wang, G. Shi, S. D. Lei, Z. Lin, X. L. Zou, G. L. Ye, R. Vajtai, B. I. Yakobson, H. Terrones, M. Terrones, B. K. Tay, J. Lou, S. T. Pantelides, Z. Liu, W. Zhou, and P. M. Ajayan,  Nature Materials 13 (12), 1135 (2014).
8   C. Thomazeau, C. Geantet, M. Lacroix, V. Harle, S. Benazeth, C. Marhic, and M. Danot,  Journal of Solid State Chemistry 160 (1), 147 (2001).
9   H. P. Komsa and A. V. Krasheninnikov,  Journal of Physical Chemistry Letters 3 (23), 3652 (2012).
10  Y. F. Chen, J. Y. Xi, D. O. Dumcenco, Z. Liu, K. Suenaga, D. Wang, Z. G. Shuai, Y. S. Huang, and L. M. Xie,  ACS Nano 7 (5), 4610 (2013).
11  D. O. Dumcenco, H. Kobayashi, Z. Liu, Y. S. Huang, and K. Suenaga,  Nature Communications 4 (2013).
12  S. J. Zheng, L. F. Sun, T. T. Yin, A. M. Dubrovkin, F. C. Liu, Z. Liu, Z. X. Shen, and H. J. Fan,  Applied Physics Letters 106 (6) (2015).
13  H. F. Liu, K. K. A. Antwi, S. Chua, and D. Z. Chi,  Nanoscale 6 (1), 624 (2014).
14  Z. Lin, M. T. Thee, A. L. Elias, S. M. Feng, C. J. Zhou, K. Fujisawa, N. Perea-Lopez, V. Carozo, H. Terrones, and M. Terrones,  Apl Materials 2 (9) (2014).
15  M. Faraji, M. Sabzali, S. Yousefzadeh, N. Sarikhani, A. Ziashahabi, M. Zirak, and A. Z. Moshfegh,  RSC Advances 5 (36), 28460 (2015).
16  J. Y. Xi, T. Q. Zhao, D. Wang, and Z. G. Shuai,  Journal of Physical Chemistry Letters 5 (2), 285 (2014).
17  L. Ma, D. N. Nath, E. W. Lee, C. H. Lee, M. Z. Yu, A. Arehart, S. Rajan, and Y. Y. Wu,  Applied Physics Letters 105 (7) (2014).
18  D. Wickramaratne, F. Zahid, and R. K. Lake,  Journal of Chemical Physics 140 (12) (2014).
19  John H.. Davies, *The Physics of Low-dimensional Semiconductors*. (Cambridge University Press, 1997).
20  A. Molina-Sanchez and L. Wirtz,  Physical Review B 84 (15) (2011).

21   A. R. Beal and W. Y. Liang, Journal of Physics C-Solid State Physics 9 (12), 2459 (1976).
22   W. Y. Liang, Journal of Physics C-Solid State Physics 6 (3), 551 (1973).
23   R. Akashi, M. Ochi, S. Borcas, R. Suzuki, Y. Tokura, Y. Iwasa, and R. Arita, e-print arXiv:cond-mat.mtrl-sci/1502.07480.
24   E. Bellotti, F. Bertazzi, and M. Goano, Journal of Applied Physics 101 (12) (2007).
25   H. Jiang, Journal of Physical Chemistry C 116 (14), 7664 (2012).
26   C. Wood and D. Jena, "*Polarization effects in semiconductors*", © Springer (2008).
27   B. K. Ridley, J. Phys. C: Solid State Phys., 15, 5899 (1982).
28   S. L. Li, K. Wakabayashi, Y. Xu, S. Nakaharai, K. Komatsu, W. W. Li, Y. F. Lin, A. Aparecido-Ferreira, and K. Tsukagoshi, Nano letters 13 (8), 3546 (2013).
29   K. Kaasbjerg, K. S. Thygesen and K. W. Jacobsen, Physical Review **B**, **85**, 115317 (2012).
30   X. Liu, J. Hu, C. Yue, N. Della Fera, Y. Ling, Z Mao and J. Wei, ACS Nano, 8 (10), 10396 (2014).
31   S. Das, H.-Y. Chen, A. V. Penumatcha and J. Appenzeller, Nano letters 13 (1), 100 (2013).